\documentclass[twocolumn,aps,prb,showpacs,amsmath,superscriptaddress,longbibliography,notitlepage]{revtex4-2}
\usepackage{amssymb,mathtools}
\usepackage[english]{babel}
\usepackage{mathrsfs}
\usepackage{graphicx}
\usepackage{float}
\usepackage{pbox}
\usepackage[caption=false]{subfig}
\usepackage{dcolumn}
\usepackage{appendix}
\usepackage{tabularx}
\usepackage{txfonts}
\usepackage[normalem]{ulem}
\usepackage{verbatim}
\usepackage{xcolor}
\usepackage{bm}
\usepackage{natbib}
\usepackage{multirow}
\usepackage{soul}

\begin{document}
\title{Twisted topology and Bipolar Non-Hermitian Skin Effect  induced by long-range asymmetric coupling }
\author{S. M. Rafi-Ul-Islam}
\email{rafiul.islam@u.nus.edu}
\affiliation{Department of Electrical and Computer Engineering, National University of Singapore, Singapore 117583, Republic of Singapore}
\author{Zhuo Bin Siu}
\email{elesiuz@nus.edu.sg}
\affiliation{Department of Electrical and Computer Engineering, National University of Singapore, Singapore 117583, Republic of Singapore}
\author{Haydar Sahin}
\email{sahinhaydar@u.nus.edu}
\affiliation{Department of Electrical and Computer Engineering, National University of Singapore, Singapore 117583, Republic of Singapore}
\affiliation{Institute of High Performance Computing, A*STAR, Singapore 138632, Republic of Singapore}
\author{Md. Saddam Hossain Razo}
\email{shrazo@u.nus.edu}
\affiliation{Department of Electrical and Computer Engineering, National University of Singapore, Singapore 117583, Republic of Singapore}
\author{Mansoor B. A. Jalil}
\email{elembaj@nus.edu.sg}
\affiliation{Department of Electrical and Computer Engineering, National University of Singapore, Singapore 117583, Republic of Singapore}
\begin{abstract}
We investigate  the twisted topology of the complex eigenspectrum of a one-dimensional non-Hermitian system under the influence of long-range unidirectional coupling. Unlike the complex energy spectrum of the conventional Hatano-Nelson chain, which takes the form of a single loop with a topological winding index of a definite sign, the introduction of long-range unidirectional hopping results in the creation of multiple twisted loops. These twisted loops exhibit opposite signs of the topological winding index, which correlate to alternating clockwise and anticlockwise energy windings. The simultaneous presence of both signs of the winding index translates into a bipolar non-Hermitian skin effect (NHSE), which challenges the conventional wisdom that the NHSE localization is dependent on the direction of the  dominant nearest-neighbor interactions. In this bipolar NHSE, the exponents of the complex energy eigenvectors corresponding to clockwise and anti-clockwise windings, lie inside and outside of the complex unit circle, respectively. Interestingly, at the intersections of oppositely oriented energy loops where the sign of the topological winding index flips, the energy becomes real-valued, leading to a suppression of the NHSE. This marks the emergence of Bloch-like contact points, where both the bipolar NHSE and the traditional NHSE vanish. Based on the non-Hermitian model we provide analytical insights into the effects of long-range unidirectional coupling to the winding topology of its complex energy spectra and their broader implications for the field of condensed matter physics.
\end{abstract}
\maketitle
\section{Introduction}
In recent years, topological phases \cite{senthil2015symmetry,rafi2022valley,fujita2011gauge,tan2020physics,tan2012introduction,ma2019topological,rafi2023conductance,hafezi2013imaging,rafi2020topoelectrical} and non-Hermitian physics \cite{ashida2020non,rafi2021topological,yao2018edge,leykam2017edge,rafi2021non,gong2018topological} have become rapidly growing fields in condensed matter physics \cite{bergholtz2021exceptional,wang2019non,martinez2018topological} and attracted the attention of researchers owing to their unique properties \cite{okuma2022non,ghorashi2021non} and potential applications in various areas \cite{bao2021fundamental,mcdonald2020exponentially,chen2021quantum,budich2020non}. 

Fascinatingly, the investigation of non-Hermitian systems, rooted in the early days of quantum mechanics, has flourished into a dynamic research field, marked by notable advancements over the past few decades \cite{bender1998real,hatano1996localization,hatano1997vortex,hatano1998non,rudner2009topological,jazaeri2001localization}. These non-Hermitian systems manifest novel and exotic physical phenomena absent in their Hermitian counterparts. In essence, non-Hermiticity in these systems arises from coupling asymmetry or the presence of gain or loss terms at the onsite level \cite{okuma2020topological,lin2023topological,li2022gain}, contributing to a rich tapestry of diverse and intriguing behaviors. The coupling asymmetry in non-Hermitian lattice models lies at the origin of many non-Hermitian phenomena ranging from the non-hermitian skin effect (NHSE)\cite{okuma2020topological,rafi2022critical,li2020critical,song2019non,rafi2022unconventional,rafi2022interfacial},  and exceptional points \cite{kawabata2019classification,minganti2019quantum}, which may be utilized in ultra-sensitive sensing \cite{budich2020non,hokmabadi2019non,de2022non}, exponential signal enhancement \cite{mcdonald2020exponentially,bao2022exponentially,koch2022quantum}, and unidirectional transport \cite{longhi2015non,longhi2015robust,du2020controllable}.

However, the effect of long-range coupling and its asymmetry on non-Hermitian systems have not been fully analyzed. Incorporating unidirectional long-range coupling into these platforms could open up a new realm of opportunities to investigate various intriguing aspects of non-Hermitian band topology. Recently, it was found that such long-range coupling asymmetry introduces peculiar phenomena associated with non-Hermiticity such as Type-II corner modes \cite{rafi2022type,chen2022observation,yang2022observation}, enhancement of topological boundary modes \cite{viyuela2018chiral,wang2022scaling} and  complex energy braiding \cite{wang2021topological,li2022topological}. Furthermore, the inclusion of asymmetric long-range coupling can lead to dramatic changes in the complex energy eigenspectra and their corresponding topology. The interplay between unidirectional long-range coupling and non-Hermiticity has also been studied in various platforms such as photonics \cite{zhu2020photonic}, metamaterial \cite{thevamaran2019asymmetric,wen2020asymmetric}, optics \cite{kominis2017stability,ivars2022optical}, condensed matter \cite{rafi2022interfacial,jin2019bulk}, and topolectrical circuit systems \cite{rafi2022unconventional,sahin2023impedance,rafi2023valley,helbig2020generalized,rafi2022system,rafi2020realization,hofmann2020reciprocal,rafi2020anti,zhang2023anomalous}. In photonic systems, asymmetric long-range coupling can be achieved by using waveguides with asymmetric coupling coefficients \cite{bell2017spectral}, while in metamaterials, it is implemented using asymmetrical split-ring resonators  \cite{decker2011retarded,roy2012coupling}. Likewise, in condensed matter systems, such coupling can be modelled via asymmetric long-range electronic hopping between atoms \cite{longhi2021non,kawabata2020higher}, while in topolectrical circuits, it can be realized through asymmetric circuit components such as operational amplifiers \cite{rafi2021non,helbig2020generalized,imhof2018topolectrical}.
 
In this work, we investigate the creation of complex energy spectra with multiple and arbitrary number of twisted loops in the presence of  long-range unidirectional hopping in a Hatano-Nelson (HN) \cite{li2020critical,suthar2022non,schindler2021dislocation} chain. The complex energy spectrum of a finite HN chain has a non-trivial topology \cite{yang2022non,ghatak2019new} that can be characterized by a topological invariant called the winding number \cite{leykam2017edge,yokomizo2019non,zhang2020correspondence}. In the absence of long-range coupling, the energy spectrum of the HN chain takes the form of a single closed loop and the winding number assumes a constant value with a definite sign throughout the complex energy plane enclosed by the loop. This leads to the accumulation of eigenstates at only one end of the chain depending on the direction of the dominant asymmetric coupling, which is reflected in the sign of the winding number.

This simple scenario can be dramatically modified by incorporating unidirectional long-range coupling, which modifies the trajectory of the complex eigenvalues and results in multiple twisted loops in the complex energy plane. The twisted topology translates into an unconventional NHSE configuration in which eigenstates associated with the clockwise and anti-clockwise complex energy loops are localized at opposite ends of the chain. The resulting  bipolar NHSE localization overturns the conventional expectation that NHSE only  occurs at one end of the NH chain corresponding to the dominant coupling direction of the asymmetric nearest-neighbor interactions. Furthermore, novel behavior was observed at the contact points between oppositely oriented energy loops, where the energy becomes real-valued. Here, the topological invariant changes sign leading to the emergence of Bloch-like contact points at which the NHSE vanishes. Our analysis of a NH model with long-range unidirectional coupling offers fresh insights into the interplay between non-Hermiticity and long-range coupling, and the resulting topology described by its complex energy eigenspectra.

\section{Effects of long-range coupling on complex energy distribution}
\begin{figure*}[htp]
    \centering
    \includegraphics[width=0.8\textwidth]{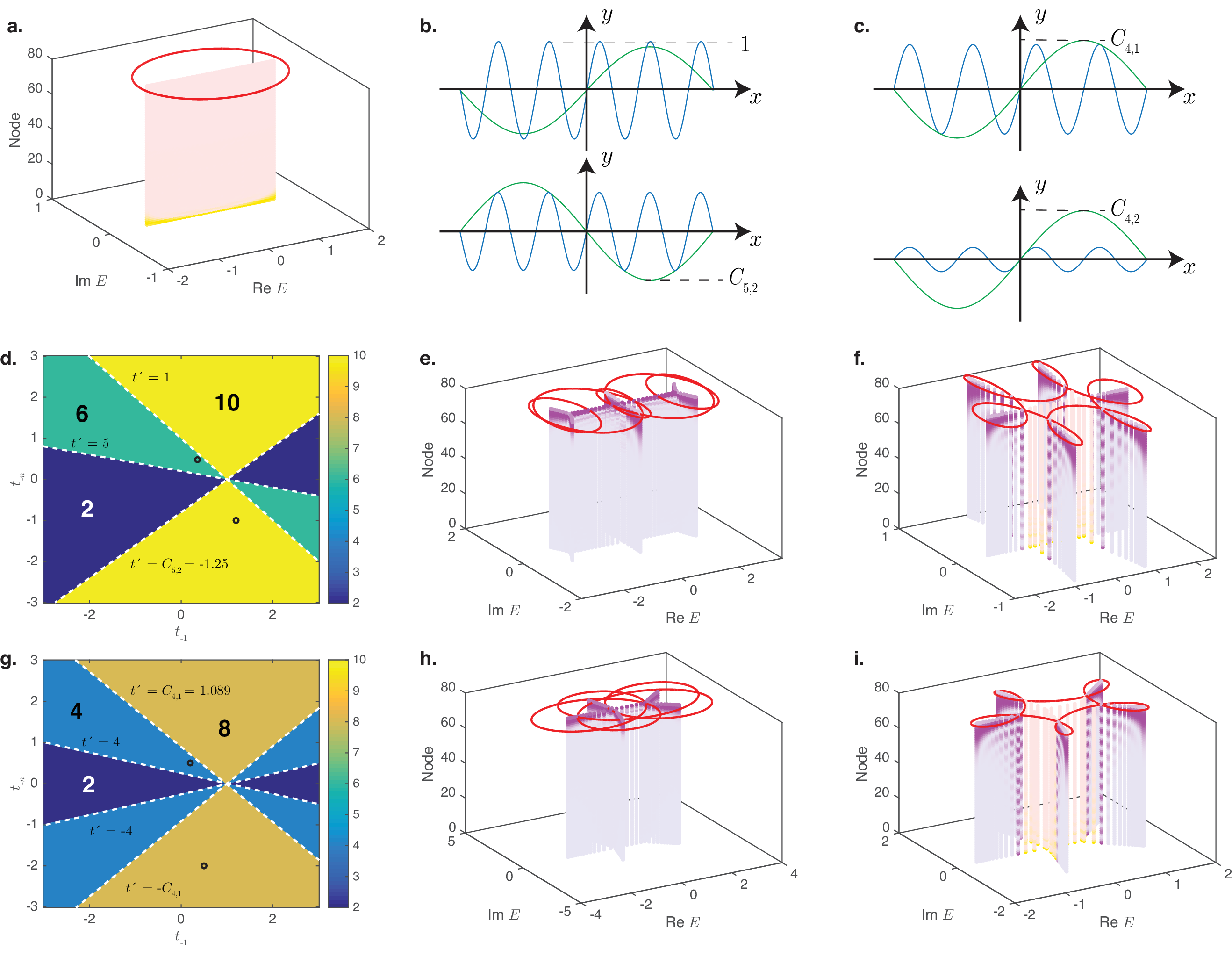}
    \caption{\textbf{ Complex eigenenergy spectra induced by long-range unidirectional coupling}
		(a) PBC spectra (thick red line), and OBC spectra and spatial probability density distribution of a conventional Hatano-Nelson chain, 80 nodes in length and without unidirectional long-range coupling ($t_{-1} = 0.4, t_{-n}=0$).(b), (c) Plots of $t'\sin x$ (green line) and $\sin(nx)$ (blue line) for (b) $n=5$, $t'=C_{5,1}=1$ (top) and $t' = C_{5,2} = 1.250$ (bottom), and (c) $n=4$, $t'= C_{4,1} = 1.089$ and $t'=C_{4,2} = 4$. (d) Plot of the number of times that the PBC eigenenergy spectrum curve intersects the real energy axis as a function of $t_{-1}$ and $t_{-n}$ for $n=5$. The boundaries between regions with different number of intersection points are demarcated by the white dotted lines and the corresponding values of $t'$ indicated.   (e) and (f) Exemplary PBC eigenenergy spectra and OBC eigenenergy spectra and spatial density distribution for $n=5$ with (e) $t_{-1}=1.2$, $t_{-5}=-1$, for which the eigenstates are all localized towards the right, and (f) $t_{-1}=0.4$, $t_{-5}=0.5$, which exhibits bipolar NHSE. The $(t_{-1},t_{-5})$ values for (e) and (f) are marked by dots on (d). (g)-(i). Corresponding plots for $n=4$ with (h) $t_{-1}=0.5$, $t_{-4}=-2$ and (i) $t_{-1}=0.2$, $t_{-4}=-0.5$.  The darker dots in the spatial probability density distributions represent high density values. The states that are localized towards the left (smaller node numbers) are denoted by green dots and those localized towards the right (larger node numbers) by blue dots. $t_{1}=1$ for all the plots in the figure. }
    \label{gFig1} 
\end{figure*}

We consider a periodic one-band system with non-reciprocal nearest-neighbor couplings $t_1 \neq t_{-1}$, i.e., the Hatano-Nelson (HN) model \cite{hatano1996localization,hatano1998non,lin2023topological}, whose eigenenergy is given by 
\begin{align}
	E_{\mathrm{HN}} &= t_1 \exp(i k) + t_{-1} \exp(-i k) \nonumber \\
	& = (t_1 + t_{-1})\cos(k) + i (t_1 - t_{-1})\sin (k).
	\label{EqHam1}
\end{align}
The locus of $E_{\mathrm{HN}} $ takes the form of an ellipse in the complex energy plane with the axis lengths of $|t_1 + t_{-1}|$ and $|t_1 - t_{-1}|$ (Fig. \ref{gFig1}a). The ellipse cuts across the real energy axis twice under periodic boundary condition (PBC).

The energy eigenspectrum exhibits distinct distributions under open boundary conditions (OBC) from the PBC eigenspectrum owing to the asymmetrical coupling between lattice sites. The tight-binding Hamiltonian corresponding to Eq. \eqref{EqHam1} for a finite HN chain that extends from $x=1$ to $x=N$ with OBC is given by 
\begin{align}
    H_{\mathrm{HN; OBC}} = \sum_{x=1}^{N-1} |x\rangle t_1 \langle x+1| + |x+1 \rangle  t_{-1} \langle x| \label{eqr1}
\end{align}
where $|x\rangle$ and $\langle x |$ are ket and bra vectors representing the basis states at site $x$.

We derive the eigenenergy spectrum of Eq. \eqref{eqr1} using the imaginary gauge approach in Appendix A and the generalized Brillouin zone (GBZ) approach in Appendix B, and show that in the eigenenergies $E$ lies on the real energy axis with  $|E| < 2\sqrt{t_1t_{-1}}$ (Fig. \ref{gFig1}a). The marked disparity between the OBC and PBC eigenspectra (line vs. ellipse on the complex energy plane) heralds the breakdown of the conventional bulk boundary correspondence (BBC) in a non-Hermitian system. Specifically, in non-Hermitian systems with coupling asymmetry, the eigenstates under OBC become localized near a single edge of the system in the non-Hermitian skin effect (NHSE). As discussed in detail in  Appendix A, the eigenstate localization direction depends on the relative magnitudes of $t_1$ and $t_{-1}$ - the NHSE localization occurs at the left (right) edge when $\ln |t_{1}/t_{-1}|$ is positive (negative). 

We now consider the introduction of a long-range unidirectional coupling $t_{-n}\exp(-ink)$ along the left direction  a distance of $n$ nodes away. The eigenenergy $E$ for a periodic system now takes the form of 
\begin{align}
	E =& t_1 \exp(i k) + t_{-1} \exp(-i k) + t_{-n} \exp(-i nk) \nonumber \\
	=& (t_1 + t_{-1})\cos (k) + t_{-n}\cos(nk) \nonumber \\
	&+ i \left( (t_1 - t_{-1})\sin(k)  - t_{-n} \sin(nk) \right). \label{eqE}
\end{align}   

The locus of the eigenenergy now intersects the real energy axis more than the two times it does in the conventional HN model. The intersection between the eigenenergy locus and the real energy axis is governed by the following equation:
\begin{align}
	& (t_1 - t_{-1})\sin(k)  - t_{-n} \sin(nk) = 0 \\
	\Rightarrow& \sin(nk) = \frac{t_{1}-t_{-1}}{t_{-n}}\sin(k)  \label{eqAuxCutCond} 
\end{align}
Introducing $t' \equiv  \frac{t_{1}-t_{-1}}{t_{-n}}$, the number of real solutions for of $k $ that satisfies Eq. \eqref{eqAuxCutCond} in the range of $ -\pi < k < \pi$, depends on the value of $t'$. For illustration,  $\sin(k)$ is plotted together with $t'\sin(nk)$ for an odd value of $n=5$ and even value of $n=4$ (see Fig.  \ref{gFig1}b and  \ref{gFig1}c, respectively). It can be seen that the maximum number of times the $t'\sin(nk)$ curve can intersect the $\sin(k)$ curve is $2n$ times regardless of whether $n$ is even or odd. Moreover, the $\sin(k)$ curve always touches the $t'\sin(nk)$ curve at $k=0, \pi$ regardless of the value of $t'$ or $n$. 

For odd $n$, the $t'\sin(nk)$ curve intersects the $\sin(k)$ curve the maximum number of $2n$ times when $0<t'<1$. However, when $t'$ exceeds 1 slightly, the maximum value of $|t'\sin(k)|$ at $k=\pm \pi/2$ now exceeds the maximum value of $\sin(k)$ at  $\sin(\pm \pi/2) = 1$.  This results in the number of intersections between the two curves decreasing by four since the $t'\sin(nk)$ curve now no longer intersects the $\sin(k)$ curve at $k=\pi/2 \pm \delta k$ and at $k=-\pi/2 \pm \delta k$.  We denote this first critical value of $t'$ which results in the lowering of the number of intersection points  as $C_{n, 1}=1$. When $t'$ is increased further to beyond $t'=n$, the $t'\sin(nk)$ curve no longer intersects the $\sin(k)$ curve near $k=\pm \pi$ and both sides of $k=0$. This results in a further decrease by four in the number of intersection points. Moreover, when $t'$ is negative, and its  magnitude  is increased, we approach a second critical value of $t'=C_{n, 2}$ beyond which the $t'\sin(k)$ curve no longer intersects the $\sin(nk)$ curve near $k = \pm \pi (\frac{1}{2} \pm \frac{1}{2n})$ (see lower plot of Fig. \ref{gFig1}b), and the number of intersection points between the $t'\sin(k)$ and $\sin(nk)$ curves drops by further a step of eight.  For the particular case of $n = 5$ shown in Fig. \ref{gFig1}b,  $C_{5,2}$ was numerically determined to be -1.25. 

The abovementioned trends for the number of intersection points between the $\sin(k)$ and $t'\sin(nk)$ curves result in the phase diagram shown in Fig. \ref{gFig1}. Notice that the number of intersection points drop in steps of 4 or 8 as explained above, and that the different phases are not symmetrically distributed about $t_{-n}=0$. In general, for larger odd values of $n$ there will be further critical values of $C_{n,m}$ at which the number of intersection points between the $t'\sin(k)$ and $\sin(nk)$ curves drops in steps of 4 or 8 from the maximum value of $2n$ down to the minimum value of 2.
  
Conversely, for even $n$, the $t'\sin(k)$ curve intersects across the $\sin(nk)$ curve for $2n$ times for small $|t'|$. As $|t'|$ increases, it reaches a critical value of $|t'| = C_{n,1}$ beyond which the $t'\sin(x)$ curve no longer intersects the $\sin(k)$ curve near $k=\pm \pi(\frac{1}{2} +  \frac{1}{2n})$. For even $n \geq 4$, as $|t'|$ increases further, there are further values of $|t'| = C_{n, m} $ beyond which the number of times the $t'\sin(k)$ curve intercepts the $\sin(k)$ curve decreases further because the  $|t'\sin(k)| > |\sin(k)|$ at increasing values of $|k-(\pm\pi/2)|$. In particular, at $|t'|$ values slightly larger $C_{n,n/2} = n$ the $t'\sin(k)$ curve no longer intersects the $\sin(k)$ curve near $k=0$ for all even $n$ (lower plot of Fig. \ref{gFig1}c ). In contrast to the case of odd $n$, the critical values of $t'$ exist in $\pm |t'|$ pairs, which gives rise to the phase diagram shown in Fig. \ref{gFig1}d, which is symmetrical about $t'_{-n}=0$. 

With these considerations on the number of intersection points and hence real solutions to the eigenenergy for the even ($n = 4$) and odd case ($n = 5$), we turn our attention to  the phase diagrams shown in Fig. \ref{gFig1}d and \ref{gFig1}g. Figs. \ref{gFig1}e, f and \ref{gFig1}h, i show the eigenenergy curves for representative examples of systems with intermediate (Fig. \ref{gFig1}e and h) and maximal (Fig. \ref{gFig1}f and i) number of intersections with the real axis, respectively. Note that these eigenenergy curves are symmetric about the real axis. This symmetry may be readily explained by the fact that since the coefficients  $t_i$s are real, $E(k) = E^*(-k)$ from Eq. \eqref{eqE}. Thus, for each value of energy $E$ corresponding to crystal momentum $k$ on the eigenenergy curve, its reflection about the real axis $E^*$ corresponding to $-k$ would also be on the eigenenergy curve. In particular, the time-reversal invariant momenta $k=0$ and $k=\pi$ lie on the real axis. 

For odd $n$, the eigenenergy curves are also symmetric about the imaginary axis, as shown in Figs. \ref{gFig1}e and f.  This symmetry is to due to the fact that $\sin(n (x\pm \pi))=-\sin(nx)$, $\cos(n (x\pm \pi))=-\cos(nx)$ for odd $n$, which implies that $E(k) = -E(k+\pi)$, i.e., for each energy $E$ corresponding to $k$ that exists on the eigenenergy curve, its reflection about the origin $-E$ corresponding to $k+\pi$ is also on the eigenenergy curve. This results in the inversion symmetry of the eigenenergy curve about $E=0$ on the complex energy plane. Correspondingly, the real part of the eigenenergy curve spans between $- (t_1 + t_{-1} + t_{-n})$ at $k=\pi$ to $(t_1 + t_{-1} + t_{-n})$ at $k=-\pi$. In contrast, for even $n$, the eigenenergy curves are not symmetric about the imaginary axis (Fig. \ref{gFig1}i, j). This is because of the fact that $\sin( n (x \pm \pi)) = \sin(nx)$ and $\cos( n ( x \pm \pi)) = \cos(nx)$ which implies that the correspondence $E(k)=-E(k+\pi)$ no longer holds for odd $n$. We find instead that the real values of $E$ now span from $- (t_1+t_{-1} - t_{-n})$ at $k=\pi$ to $t_1 + t_{-1} + t_{-n}$ at $k=0$.  

We introduce OBC to a modified HN chain of length $N$ with lattice sites located at $x=1$ to $x=N$ by setting the couplings that extend outside the extent of the chain to 0. The real-space tight-binding Hamiltonian of the chain is then given by 
\begin{equation}
    H_{\mathrm{HNL; OBC}} = \sum_{x=1}^{N-1} \left(t_1 |x+1 \rangle \langle x| + t_{-1} |x \rangle \langle x+1| \right) +  \sum_{x=1}^{N-n} \left( t_{-n}|x \rangle \langle x+n|\right).
    \label{eqr11}
\end{equation}

Remarkably, the long-range coupling can cause the OBC eigenspectrum to become complex, as depicted in Fig. \ref{gFig1}f, h, and i. This differs from the OBC eigenspectrum of a HN chain without long-range coupling, which lies completely on the real energy axis. We illustrate how a complex eigenspectrum can emerge in Appendix B. Notably, the NHSE persists as long as the eigenenergy spectra under OBC and PBC remain dissimilar. The persistence of the NHSE localization when the OBC and PBC spectra differ from each other can be intuitively understood through the following argument: As described in detail in Appendix B, the wavefunction of an OBC eigenstate $\psi(x)$ at an eigenenergy $E$ has the general form of $\sum_j c_j\beta_j^x$ where $\beta_j \equiv \exp(ik_j)$ and the $(n+1)$ $k_j$s, which are generally complex, are related to $E$ via Eq. \eqref{eqE}. The PBC eigenenergy spectrum is essentially the loci of $E$ values at which at least one of the $\beta_j$ values has a modulus of 1, which in turn corresponds to a real value of $k$. This implies that none of the $\beta$ values of the OBC eigenenergies has a modulus of 1 when the OBC eigenspectrum differs from the PBC eigenspectrum, and the wavefunction grows exponentially towards the left or the right depending on the sign of the dominant $\mathrm{ln} |\beta_j|$ component. We show in Appendix B that the OBC eigenspectrum in a sufficiently large (on the order of 5 sites for the parameter ranges here) system  is in turn given by the locus of energy values at which the two largest $|\beta_j|$s have the same moduli. 

Unlike the conventional HN system in which the eigenstates are localized near one edge of the system (Figs. \ref{gFig1}a), we see in Figs. \ref{gFig1}f and i that for some parameter ranges of $t_{-1}$ and $t_{-n}$, a peculiar bipolar NHSE appears in which the OBC eigenstates are localized around both edges of the system. (The states localized nearer the left edge are denoted by yellowish dots and located near $E=0$, while the states localized nearer the right edge are denoted by bluish circles and located further away from $E=0$. )   This unique localization of eigenstates overturns the conventional expectation that the edge at which the NHSE localization occurs is determined by the dominant nearest-neighbor asymmetric coupling direction \cite{okuma2020topological}. In the next section, we will analyze this bipolar NHSE in more detail by considering the specific example of the $n=5$ long-range non-Hermitian coupling system, whose eigenmode localization is illustrated in Figs. \ref{gFig1}e and f. 

\begin{figure*}[htp]
    \centering
    \includegraphics[width=0.8\textwidth]{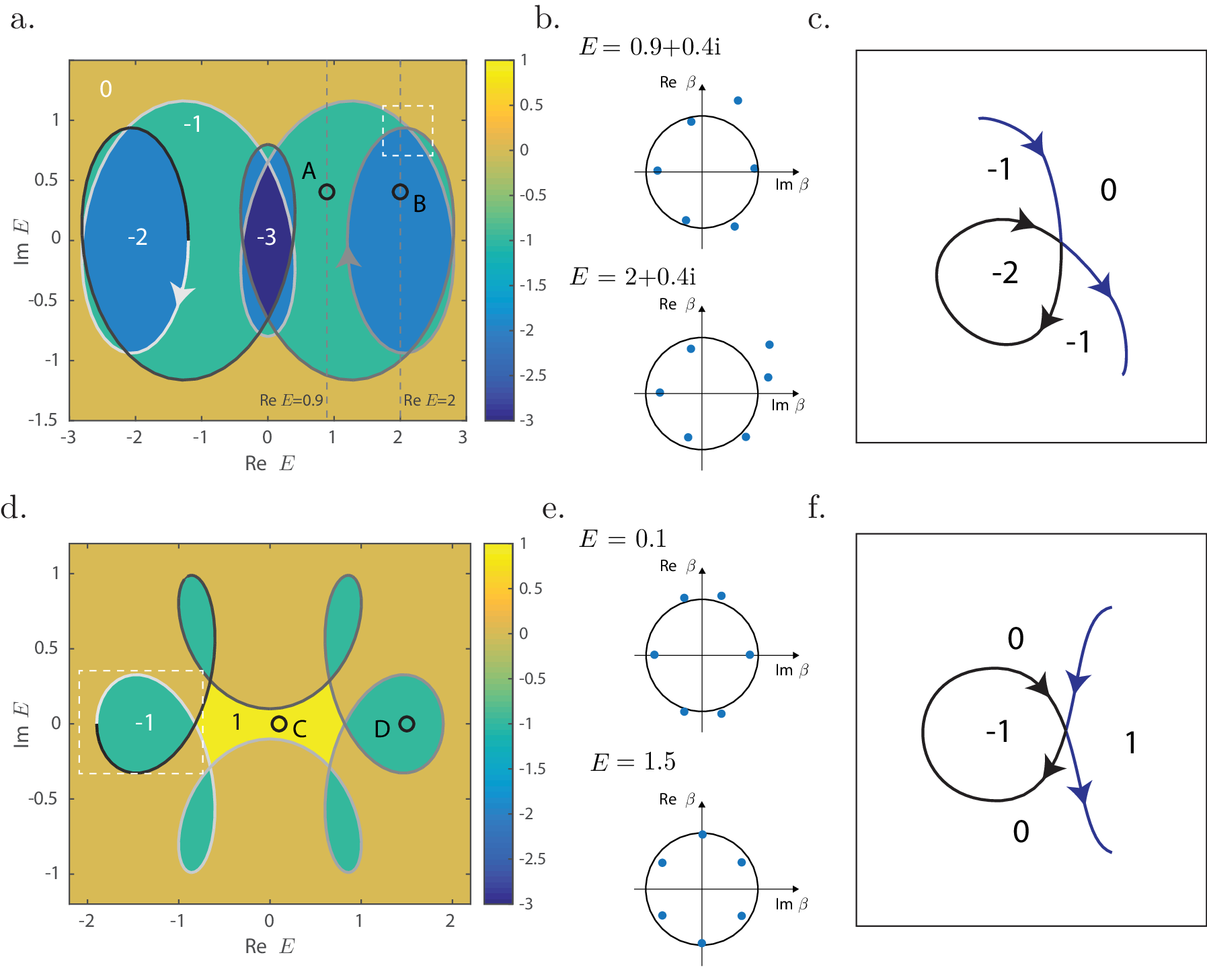}
    \caption{\textbf{ Winding number and NHSE localization :}
		(a) PBC energy spectrum and winding number distribution of $n=5$ system without bipolar NHSE corresponding to Fig. \ref{gFig1}e with $t_{-1}=1.2$ and $t_{-5}=-1$.  A lighter color on the PBC energy curve corresponds to smaller values of $k$, which range from $-\pi$ to $\pi$. The arrow on the PBC curve indicates the direction of increasing $k$. (b) The $\beta$ distribution on the complex $\beta$ plane at the complex energies of (top) $E=0.9+0.4i$ and (bottom) $E=2+0.4i$, which are indicated by the labelled circles in (a). (c) A schematic illustration of the winding number distribution around an internal loop, as exemplified by the area within the white dotted rectangle in Fig. \ref{gFig2}a.  (d) PBC energy spectrum and winding number distribution of $n=5$ system exhibiting bipolar NHSE with $t_{-1}=0.4$ and $t_{-5}=0.5$, and (e) the $\beta$ distribution at $E=0.1$ (top) and $E=1.5$ (bottom). (f) A schematic illustration of the winding number around an external loop, as exemplified by the dotted rectangle in Fig. \ref{gFig2}d.}
    \label{gFig2} 
\end{figure*}

\subsection{ Topological invariant in presence of long range coupling}
In the preceding section, we explored how the introduction of long-range coupling gives rise to skin modes at one or both boundaries. In this section, we introduce a non-Hermitian topological invariant to analyze the conventional and bipolar NHSE. The NHSE localization direction at an arbitrary reference energy $E_{\mathrm{ref}}$ can be determined from the non-Bloch winding number $\eta$, which is as
\begin{equation}
	\eta = \frac{1}{2\pi} \oint_{|\beta|=1}  \mathrm{d}\beta\ \mathrm{Arg}(E(\beta) - E_{\mathrm{ref}}).  \label{eqEta}
\end{equation}

A positive (negative) finite value of $\eta$ indicates that a semi-infinite system extending from  $-\infty$ to 0 (0 to $\infty$) would host bulk states that are NHSE-localized at the right (left) edge of the system. A positive (negative) winding number also indicates that if bulk OBC states exist within the energy region around $E_{\mathrm{ref}}$ is bounded by the PBC curve, then these OBC states will be localized at the right (left) boundaries of the system. We explain why this is so below using the concept of the GBZ. Figs. \ref{gFig2}a and c show the winding numbers at different energies on the complex energy planes for the $n=5$ systems shown in Figs. \ref{gFig1}e and f, respectively. A comparison between Fig. \ref{gFig2}a and Fig. \ref{gFig1}e, and between Fig. \ref{gFig2}d and Fig. \ref{gFig1}f shows that this correspondence between the sign of the winding numbers and the localization direction of the OBC eigenstates indeed holds.
 
Furthermore, the winding number can be visually determined from the PBC eigenenergy curve on the complex energy plane if the direction of increasing $k$ on the eigenenergy curve is known. From Eq. \eqref{eqEta}, the winding number at the reference energy $E_{\mathrm{ref}}$ is simply the number of times the eigenenergy curve winds around $E_{\mathrm{ref}}$ as $k$ is increased from $-\pi$ to $\pi$ with positive (negative) values of the winding number corresponding to counter-clockwise (clockwise) windings. As an illustration, let us consider the reference energy $E_{\mathrm{ref}}=0.9+0.4i$, indicated by the  open circle labelled ``A", in Fig. \ref{gFig2}a. Its associated winding number is -1 because the vertical line $x=0.9$ passing through the centre of this circle cuts across the eigenenergy curve twice, with the eigenenergy curve progressing from left to right with increasing $k$ above the circle and the  curve progressing from right to left below the circle, indicating a clockwise winding. Similarly, the winding number around $E_{\mathrm{ref}}=2+0.4i$, indicated by the  open circle labelled ``B", is -2 because the vertical line $x=2$ passing through  the circle cuts across the eigenenergy curve twice above the circle and twice below it. 

Besides this visual interpretation of the winding number, the winding number can also be expressed in an alternative form that gives more insight into the connection between the winding number and the NHSE localization direction. Eq. \eqref{eqEta} can be recast into 
\begin{equation}
	\eta = \frac{-i}{2\pi} \oint_{|\beta|=1} \mathrm{d}\beta\ \frac{ \partial_\beta \left(E(\beta) - E_{\mathrm{ref}}\right)}{E(\beta) - E_{\mathrm{ref}}}
\end{equation}
where $\beta \equiv \exp(ik)$. Applying the argument principle, $\eta$ is then given by $\eta(E_{\mathrm{ref}}) = Z(E_{\mathrm{ref}}) - P(E_{\mathrm{ref}})$ where $P(E_{\mathrm{ref}})$ is the number of poles of $E(\beta)-E_{\mathrm{ref}}$  while $Z(E_{\mathrm{ref}})$ is the corresponding number of zeros lying within the complex unit circle. From Eq. \eqref{eqE}, $E(\beta) - E_{\mathrm{ref}} = t_1 \beta + t_{-1} \beta^{-1} + t_{-n}  \beta^{-n} - E_{\mathrm{ref}}$.  Therefore $P=n$, while $Z(E_{\mathrm{ref}})$ is the number of $\beta$s that satisfy $E(\beta) = E_{\mathrm{ref}}$ and fall within the unit circle on the complex $\beta$ plane. Fig. \ref{gFig2}b shows the distribution of $\beta$s satisfying  $E(\beta) = E_{\mathrm{ref}}$ for $E_{\mathrm{ref}}=0.9+0.4i$ (top) and $E_{\mathrm{ref}}=2+0.4i$ (bottom). From these plots, it is evident that $Z(E_{\mathrm{ref}})$ are 4 and 3, respectively, which correspond to the respective winding numbers of $\eta = 4-5 = -1$   and $\eta = 3-5=-2$, respectively, for the two values of $ E_{\mathrm{ref}}$, noting that n = 5. The winding numbers obtained here by considering the number of poles and zeros are  in agreement with those obtained by visually counting the number of times the eigenenergy curve winds around $E_{\mathrm{ref}}$ following Eq. \ref{eqEta} and depicted in Fig. \ref{gFig2}a. Fig. \ref{gFig2}e shows corresponding examples for the two other exemplary $E_{\mathrm{ref}}$ values of 0.1 and 1.5 for the $n=4$ case.

From the above analysis, we can correlate the winding number $\eta$ to the NHSE localization direction under OBC: The fact that $\eta = Z - P = Z - n$ implies that the number of $\beta$ values that satisfy $E(\beta) = E_{\mathrm{ref}}$ and lie within the unit circle in the complex $\beta$ plane is given by $Z = n + \eta$. We show in Appendix B that for a system with $n$th-order long-range coupling to the left, the condition for  $E_{\mathrm{ref}}$ to lie on the GBZ and thus be an OBC eigenenergy in the thermodynamic limit, is $|\beta_{n+1}| = |\beta_n|$ where $|\beta_1| \leq |\beta_2|  ...  \leq |\beta_{n+1}|$.  An $\eta$ value equal to $1$ indicates that all $(n+1)$ $\beta$ values lie within the unit circle at energy values `near' $E_{\mathrm{ref}}$, which implies that $\beta_n$ and $\beta_{n+1}$ would both be within the unit circle if the GBZ condition $|\beta_n|=|\beta_{n+1}|$ holds. Here, we define an arbitrary energy $E$ to be `near' $E_{\mathrm{ref}}$ if the PBC eigenenergy curve does not fall between $E$ and $E_{\mathrm{ref}}$ on the complex energy plane. We explain the reason for this definition in the next paragraph.  Thus, if an OBC eigenstate exists near $E_{\mathrm{ref}}$, then $|\beta_n|=|\beta_{n+1}| < 1$ for that eigenstate, and the eigenstate would be localized at the left edge. Conversely, if $\eta <0$ at $E_{\mathrm{ref}}$, then both  $|\beta_n|$ and $|\beta_{n+1}|$ have magnitudes larger than 1, which implies that any OBC eigenstate that exists near $E_{\mathrm{ref}}$ would be localized at the right edge.  Finally, an  $\eta$ value of exactly 0 would indicate that exactly $n$ of the smaller $\beta$ values near $E_{\mathrm{ref}}$, including $\beta_n$, lie strictly inside the unit circle while $\beta_{n+1}$ lies outside. This implies that the condition $|\beta_n|=|\beta_{n+1}|$ for an OBC eigenstate to exist cannot be satisfied, so there are no OBC eigenstates near $E_{\mathrm{ref}}$.      

One corollary of $\eta= Z - P $, i.e.,  the difference between the number of poles and zeros of $E(\beta)-E_{\mathrm{ref}}$, is that changes in the winding number as $E_{\mathrm{ref}}$ is varied across the complex plane, would always entail a crossing of the PBC eigenenergy curve. This is because any change in the winding numbers  will involve the transit of at least one of the $\beta$ roots of the $E(\beta) = E_{\mathrm{ref}}$ equation across the unit circle boundary in the complex $\beta$ plane. Thus, during this transition, at least one of  the $\beta$ values must lie exactly on the complex unit circle at some point.  When $\beta=\beta'$ lies exactly on the complex unit circle, its corresponding $k=-i\ln(\beta')$ value is real, which in turn implies that $E(\beta')$ lies on the PBC energy spectrum. Consequently, the winding number changes by 1 across a simple linear section of the PBC spectrum,  such as depicted by  the blue line segment in Fig. \ref{gFig2}c. We can thus conclude  that the NHSE must exist within any non-Hermitian system in which the eigenenergy curve encloses a finite area . This is because the winding number within the interior regions bounded by the eigenenergy curve has a finite value, since it differs by 1 from the winding number (which is 0) in the region outside the curve that extends to infinity. Conversely, there will be no NHSE when the eigenenergy curve takes the form of an open curve which does not enclose any finite area in the complex energy plane.

\subsection{Mechanism of complex energy loop crossing and its correspondence with NHSE}
In the preceding subsection, we have examined the impact of the interplay between complex energy and long-range coupling on the sign and magnitude of the winding number, which in turn affects the localization of the NHSE. Intriguingly, the presence of long-range coupling can alter the complex energy spectra and lead to intersecting eigenenergy curves and various loop configurations. In this subsection, we discuss how long-range coupling also induces and modifies the complex eigenenergy loop crossings and discuss their correspondence with the winding number and NHSE.

As an illustration, Fig. \ref{gFig2}c shows a schematic representation of an internal loop demarcated by the white dotted rectangle in Fig. \ref{gFig2}a in which the eigenenergy curve intersects itself in a crossing that resembles an ``X" shape to form an inner loop enclosed within a larger outer loop in  the complex energy plane area bounded by the curve. We shall refer to such self-crossings as X crossings for conciseness henceforth. The intersection point in an X crossing corresponds to having two values of $\beta$ lying on the unit circle on the complex  plane. Consequently, when one traverses across the two opposite sides of the X crossing, the winding number must change by either $\pm 2$ corresponding to having both $\beta$ values moving into or out of the unit circle, or remain unchanged which corresponds to one $\beta$ value moving into the unit circle while the other $\beta$ moves out of it.

By considering the direction of increasing $k$ on the eigenenergy curve denoted by the arrowheads in Fig. \ref{gFig2}c, it can be seen that the latter (i.e. $0$ change in the winding number) corresponds to the two quadrants of the X-crossing bounded by arrowheads pointing both towards or away from the crossing point (i.e., the upper and lower quadrants of the X-crossing), and the former (i.e. $\pm 2$ change in the winding number) involves the quadrants bounded by one arrowhead pointing towards and the other  pointing away from the crossing point (the upper and lower quadrants of the X-crossing).  Additonally, as shown in Fig. \ref{gFig2}c, the winding number within the internal loop has the same sign as that the larger outer loop bounded by the eigenenergy curve.

We now consider the self-intersection  of the PBC eigenenergy curve to form an external loop that protrudes into the region exterior to the eigenenergy curve. (Note that ``intersection" here refers to the eigenenergy curve intersecting itself, while the related term ``crossing" refers to the eigenenergy curve crossing the real axis.). One example is demarcated by  the dotted white rectangle in Fig. \ref{gFig2}d, whose schematic representation is shown in Fig. \ref{gFig2}f. Because an external loop is directly adjacent to the region exterior to the PBC curve, its winding number is necessarily $\pm 1$ (since the winding number associated with the exterior region is $0$). Furthermore, by considering the direction of increasing $k$ on the eigenenergy curve, it can be seen that the winding number changes by $\pm 2$ as we traverse between the two quadrants of the X-crossing that link the external loop to the interior of the region bounded by the eigenenergy curve. This results in the former having a winding number of opposite sign to that in the latter. The different signs of the winding number translates to the phenomenon of bipolar NHSE in which the OBC eigenstates are localized along both edges of the finite system.
\begin{figure*}[htp]
    \centering
    \includegraphics[width=0.75\textwidth]{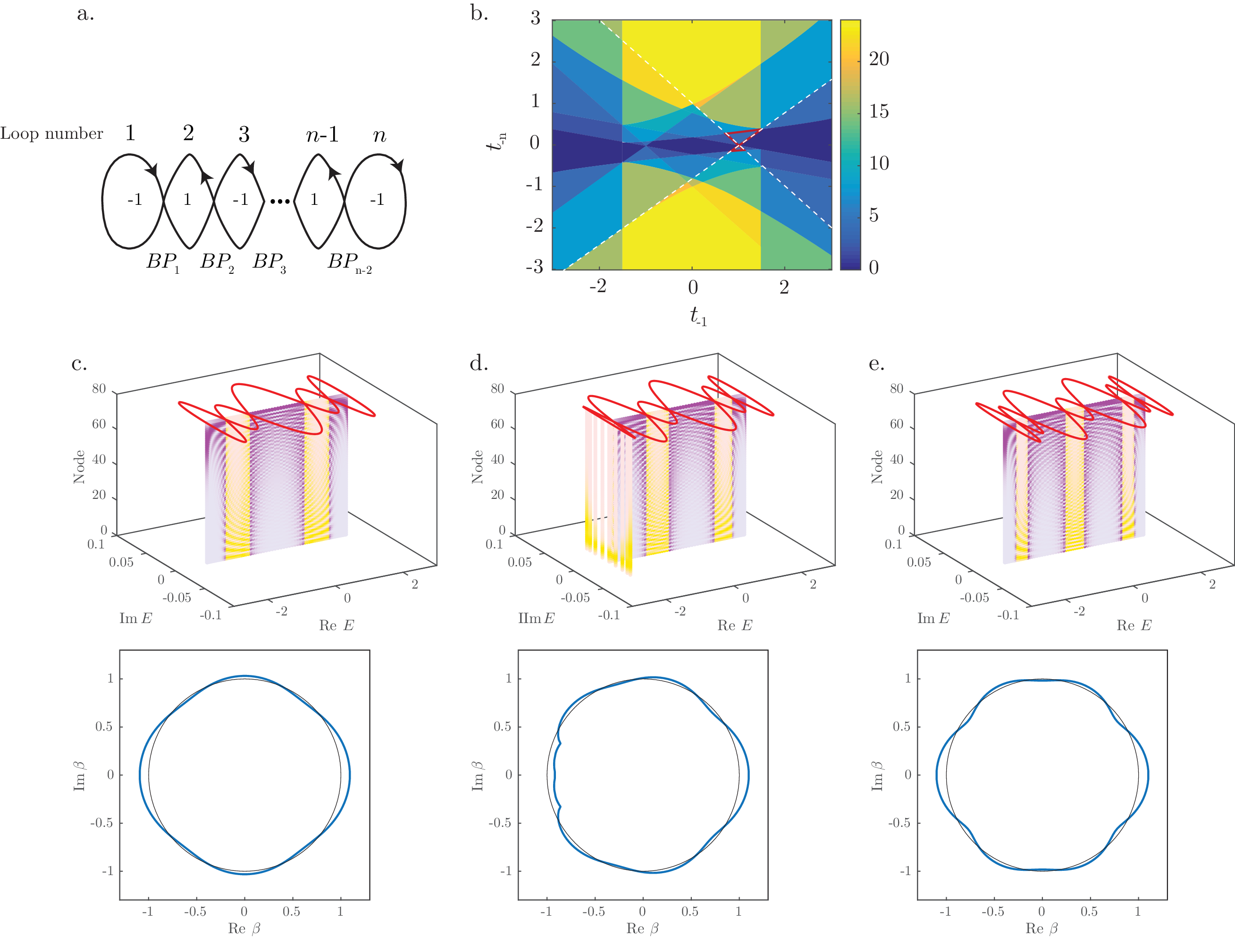}
    \caption{\textbf{ Complex energy loops with alternating signs of winding numbers}
		(a) Schematic representation of system with $n$ complex energy loops with alternating signs of winding numbers and $(n-1)$ X crossings denoted as BP$_1$ to BP$_{n-1}$. (b) Number of self-intersections in complex eigenenergy spectrum at $n=5$ as function of $t_{-1}$ and $t_{-5}$. The area demarcated by the solid red lines represents the region in which the eigenenergy spectrum takes the form of alternating loops shown in Fig. \ref{gFig3}a. The dotted white line demarcates the region on the $t_{-1}$--$t_{-5}$ plane in which the eigenenergy distribution crosses the real energy axis 10 times. (c) - (e) (Top) PBC eigenenergy spectrum, and OBC eigenenergy spectrum and spatial density distribution, and (bottom) $\beta$-plane GBZ, of (c) $n=5$, (d) $n=6$, and (e) $n=7$ systems with $t_{1}=1$, $t_{-1}=1.05$, and $t_{-n}=0.05$. }
    \label{gFig3} 
\end{figure*}

Having established the correlation between the external eigenenergy loops and the emergence of the bipolar NHSE, we will proceed to show that the number of switches between the NHSE localization at the two ends of the chain can be engineered by tuning the system parameters so to increase the number of X-crossings. We have earlier explained how the eigenenergy of an $n$-th order long-range coupled system described by Eq. \ref{eqE} can cross the imaginary axis up to a maximum of $2n$ times. Hence, a large number of alternations between NHSE localizations can be achieved by tuning the system parameters such that all the $2(n- 1)$ real-axis eigenenergy crossings (disregarding the two crossings at the extreme ends of the eigenenergy curve) take the form of X crossings.  In other words, we have $(n - 1)$  X-crossings (n.b. each X-crossing corresponds to $2$ crossings of the real axis by the two degenerate states at each X-point) resulting in $n$ loops, as shown schematically in Fig. \ref{gFig3}a. Such a configuration corresponds to the maximum number of times the complex energy loop crosses the real energy axis and the minimal number that the loop intersects itself  where all the self-intersections occur at the crossings.  Fig. \ref{gFig3}b plots the number of self-intersections for a $n=5$ long-range coupled system as a function of the coupling parameters $t_{-n}$ and $t_{-1}$.  The intersection between the region with the maximal number of real energy axis crossings demarcated by the dotted white lines (refer also to Fig. \ref{gFig1}d) and the region with the minimal number of self-intersections is denoted by the solid white lines marks. Within this intersection region, we have the desired eigenenergy spectrum which takes the form of consecutive loops along the real axis (schematically drawn in Fig. \ref{gFig3}a). From Fig. \ref{gFig3}b, we see that such a configuration exists around the vicinity of $t_1=1$, $t_{-n}=0$ (although not exactly at $t_{-n}=0$, which would correspond to a conventional HN system.)  

\subsection{Formation of series of twisted complex energy loops with  opposite windings and Bloch points}
To illustrate the concept of series of twisted complex energy loops with opposite winding index (shown schematically in Fig. \ref{gFig3}a), we consider various order of unidirectional long range coupling. With the optimal choice of the coupling parameters (e.g. corresponding to the triangular areas bounded by the white dotted lines in Fig. \ref{gFig3}b for the case $n =5$), we depict the respective examples for $n = 5$ to $7$ for which the eigenenergy spectrum takes the form of $n$ consecutive loops. As we traverse each X-crossing, the winding direction (clockwise / anticlockwise) of the eigenenergy curve flips. This leads to a change in the sign of the winding number and the flipping of the NHSE localization direction of the OBC states between successive loops. The existence of bipolar NHSE localization implies that the GBZ would contain segments that lie both within and outside the unit circle on the complex $\beta$ plane, as verified  in the bottom plots of Figs. \ref{gFig3}b  to \ref{gFig3}e.   

Let us consider the Bloch-like points that lie exactly on the complex unit circle  in Figs. \ref{gFig3}b to \ref{gFig3}e, which correspond to the $k$ values at which the eigenenergy curve intersects the real axis and forms the X-crossings. Interestingly, these OBC eigenstates that lie exactly at each X crossing are not localized at either of the edges. This absence of localization can be explained in terms of the following: Each internal intersection point at a X crossing on the complex energy plane represents a transition between a region with $\eta=1$ which possesses $(n+1)$ $\beta$ values lying within the unit circle, 
and a region with $\eta=-1$ which possesses  $(n-1)$ $\beta$ values within the unit circle (and hence two $\beta$ values outside it). Thus, each $X$ crossing corresponds to an energy value where two $\beta$ states lie exactly on the complex unit circle as one state moves from outside the unit circle to inside, and the other state moves in the opposite direction. At this intersection point, there are $(n-2) $  $\beta$  values inside the complex unit circle and one $\beta$ value outside. This means that the $n$ and $(n+1)$th largest $|\beta|$ values are  both on the complex unit circle. In other words, the $\beta$-plane GBZ on the X-crossing falls on the complex unit circle. Thus, the NHSE localization length is 0, which results in the absence of NHSE localization.  

\section{Conclusion}
In conclusion, we have analyzed the generalized Hatano-Nelson chain which incorporates the effects of asymmetric long-range coupling of arbitrary orders. The introduction of unidirectional long-range coupling results in multiple twisting topology of its PBC eigenspectra in complex energy space. The number of eigenenergy loops equates to the order of coupling. Furthermore, we showed the close correlation between the loop topology and the NHSE localization of the eigenstates of the system. The loop topology can be engineered by tuning the long-range coupling parameters to exhibit maximum number of external loops of alternating winding numbers. This leads to the phenomenon of bipolar NHSE, where the position of the NHSE localization is determined by the sign of the winding index rather than the direction of coupling asymmetry according to the conventional understanding. Interestingly, the crossing points of the eigenenergy loops are characterized by real energy Bloch-like points, at which the NHSE localization vanishes. We provide the underlying theoretical basis for the observed NHSE phenomena induced by the eigenenergy loop topology, as well as numerical verification of the theory. Finally, our results provide a flexible and accessible method to modulate the complex energy spectra of a long-range coupled non-Hermitian system, thereby realizing multiple twisted eigenenergy loop topology and inducing the novel NHSE localization as described above.
\subsubsection*{Acknowledgements}
This work is supported by the Ministry of Education (MOE) Tier-II Grant MOE-T2EP50121-0014 (NUS Grant No. A-8000086-01-00), and MOE Tier-I FRC Grant (NUS Grant No. A-8000195-01-00).

\section*{Appendix A: Localization of eigenstates in Hatano-Nelson model: imaginary gauge transformation approach}
The imaginary gauge transformation (similarity transformation) is a historically important approach in understanding the exponential localization and eigenenergy spectrum of the HN chain without long-range coupling. In this section, we apply this approach on the finite chain represented by the Hamiltonian Eq. \eqref{eqr1}. An imaginary gauge transformation 
\begin{align}
    |x\rangle &\rightarrow  |x \rangle \rangle e^{-g x} \label{kketx},\\
    \langle x| &\rightarrow e^{g x}\langle \langle x| \label{bbrax}, \\
    g &= \ln \sqrt{\frac{t_1}{t_{-1}}} \label{g}
\end{align}
can be applied to Eq. \eqref{eqr1} to convert it into the equivalent Hermitian Hamiltonian
\begin{equation}
    H_{\mathrm{OBC}} = \sqrt{t_1 t_{-1}} \sum_{x=1}^{N-1} ( |x \rangle \rangle \langle \langle x+1| + |x \rangle \rangle \langle \langle x+1|),
    \label{eqr6}
\end{equation}
where the double left / right angular brackets denote the transformed basis states. 

Importantly, this transformation preserves the energy spectrum because it is a similarity transformation. Consequently, the eigenvalue spectrum of Eq. \eqref{eqr1} is identical to that of Eq. \eqref{eqr6}. This implies that the eigenspectrum remains real for any degree of non-Hermiticity as long as $\sqrt{t_1 t_{-1}}$ is real.

At the same time, the introduction of the imaginary gauge potential induces a position-dependent scaling of the eigenfunctions. From Eq. \eqref{bbrax}, we read off that the spatial wavefunction of the eigenstate $|\psi\rangle$ in the basis of Eq. \eqref{eqr1}, $\langle x|\psi(x)\rangle$,  is related to that in the basis of Eq. \eqref{eqr6}, $\langle\langle x|\psi(x)\rangle$ by

\begin{equation}
    \langle x | \psi \rangle = e^{g x} \langle \langle x | \psi \rangle.
    \label{eqr7}
\end{equation}

The $e^{gx}$ term on the right of Eq. \eqref{eqr7} results in an exponential growth or decrease of the right wavefunction depending on the sign of $g$. All the eigenstates therefore become localized at one of the chain edges under OBC when the coupling is asymmetrical ($t_1 \neq t_{-1}$). Eq. \eqref{eqr6} is essentially the lattice version of Hamiltonian one-dimensional free electron gas in an infinite potential well, which has the well-known solution $\langle \langle x|\psi\rangle = e^{i k x} - e^{-i k x}$ where $k = (2 n \pi/(N+1))$, $n=1, ..., N$ for a chain that extends from $x=1$ to $x=N$. From the Bloch counterpart of Eq. \eqref{eqr6}, $H(k)=2\sqrt{t_1t_{-1}}\cos(k)$, it can be deduced that the OBC eigenenergies of Eq. \eqref{eqr7} lie on the real line $|E| \leq 2\sqrt{t_1t_{-1}}$.

The relationship between $\langle x|$ and $\langle \langle x|$ then implies that the wavefunction of an OBC eigenstate $\psi_{HN}(x)\equiv \langle n|\psi_{\mathrm{HN}}$ in the basis of Eq. \eqref{eqr1} is explicitly given by

\begin{equation}
    \psi_{\mathrm{HN}}(x) = \left(\frac{t_{-1}}{t_1}\right)^{x/2} (e^{ikx} - e^{-ikx}).
    \label{eqr10}
\end{equation}
 
A notable observation from Eq. \eqref{eqr10} is that the magnitude of the wavefunction is directly proportional to $\left(\frac{t_{-1}}{t_1}\right)^{N/2}$. This dependence has a significant implication: when $|t_{-1}| > |t_1|$, the eigenstates become exponentially localized near the right edge of the chain at $x = N$. Conversely, if $|t_{-1}| < |t_1|$, the eigenstates exhibit exponential localization near the left edge of the chain at $x = 1$. 

Although the imaginary gauge transformation described above sheds light on the effects of asymmetrical coupling and the spatial distribution of the wavefunction in simple non-Hermitian systems like the non-reciprocal Hatano-Nelson system above or the Su-Schrieffer-Heeger chain, it is not universally applicable to more complicated systems such as the ones with asymmetrical long-range coupling studied here. This is because there is no similarity transformation with a constant value of $g$ like that in Eq. \eqref{kketx} and \eqref{bbrax} that can be performed on a generic non-reciprocal long-range Hamiltonian to convert it into a Hermitian Hamiltonian with purely real eigenvalues. The non-existence of such similarity transformations is hinted at by the fact that unlike the HN system where $|\beta|$ has the constant value of $g$ in Eq. \eqref{g} throughout the entire GBZ, which then takes the form of a circle in the complex $\beta$ plane, the $|\beta|$ values in general vary at different points on the GBZ, as can be seen from the lower plots in Fig. \ref{gFig3}c--e. Another indication that such similarity transformations do not exist is the fact that the OBC energy spectra for these systems are complex rather than real: if a similarity transformation that converts the non-Hermitian Hamiltonian to a Hermitian one and preserves the eigenvalues exists, it would not have been possible to obtain complex eigenvalues from a Hermitian Hamiltonian. A more modern and universally applicable approach that has been commonly adopted to explain the non-Hermitian skin effect in more recent works over the past three years is the GBZ, which we explain in more detail in the next section. 

\section*{Appendix B: Generalized Brillouin zone}
As noted above, although the OBC eigenspectra of prototypical systems as the Su-Schrieffer-Heeger (SSH) and Hatano-Nelson chains with nearest-neighbor couplings are consistently real, this is not always the case for more complex models involving long-range couplings or gain/loss terms \cite{okuma2020topological,kawabata2020higher,lin2023topological,tai2023zoology,lee2019anatomy,li2020critical}.

The emergence of complex eigenenergies is not surprising because the eigenvalues of a non-Hermitian matrix, such as the Hamiltonian of a finite-length chain with asymmetrical coupling in Eq. \eqref{eqr11}, are not restricted to real values but can, in general, be complex. Fundamentally, the eigenvalue $E$ must satisfy the requirement that the Schr\"{o}dinger equation $\langle x|H|\psi\rangle = \langle x|\psi\rangle E$ is satisfied by the eigenstate of a Hamiltonian of a finite-length chain (i.e., under OBC) at all the lattice sites lying within the extent of the chain, i.e., $x=1,...,N$. 

Consider the Hamiltonian Eq. \eqref{eqr11}. For a lattice site $x$ that lies within the interior in the chain for which all the sites it is coupled to by $t_{-1}$, $t_{1}$, and $t_{-n}$ lie within the chain, i.e., $n < x < N-1$, the Schr\"{o}dinger equation at $x$ reads 
\begin{equation}
    t_1 \psi(x+1) + t_{-1} \psi(x-1) + t_{-n} \psi(x-n) = E \psi(x). \label{pbcH} 
\end{equation}
This is the same equation that is obeyed at any lattice site inside an \textit{infinitely} long chain. In non-Hermitian systems, the  Bloch theorem for Hermitian system, which states that the wavefunction of a periodic system with a unit cell containing a single lattice point has the form of $\exp(ikx)$, is extended so that $k$ is no longer limited to real values but can, in general, be complex (see, for example, \cite{PRL123_066404}). It is conventional to introduce $\beta\equiv\exp(ik)$. Writing  $\psi(x)=\beta^x$ in Eq. \eqref{pbcH} gives 
\begin{equation}
    t_1 \beta^{n+1} + t_{-1} \beta^{n-1} + t_{-n} - E\beta^n = 0,  \label{Epbc} 
\end{equation}
which is an $(n+1)$th-order polynomial in $\beta$. For a given $E$, Eq. \eqref{Epbc} has $n+1$ solutions for $\beta$, which we label as $\beta_1, \beta_2, ..., \beta_{n+1}$ where $|\beta_1|\leq|\beta_2|\leq ... \leq|\beta_{n+1}|$. Eq. \eqref{pbcH} is then satisfied by any linear combination of the $n+1$ $\beta$ values
\begin{equation}
    \psi(x)=\sum^{n+1}_{j=1} \beta_j^x c_j \label{psix}
\end{equation}
where the $n+1$ $c_j$s are position-independent constant coefficients. In particular, $\psi(x)$ in Eq. \eqref{psix} is also an eigenstate of the OBC Hamiltonian Eq. \eqref{eqr11} when appropriate boundary conditions are applied as follows: We note that Eq. \eqref{pbcH} will also hold for an eigenstate of Eq. \eqref{eqr11} at $1 \leq x \leq n$ and $x=N$ if we introduce the $n+1$ constraints that $\psi(-n+1)=...=\psi(0)=0$ and $\psi(N+1)=0$ [for example, $\langle x=1|H\rangle = E\psi(1) =  t_1 \psi(2)$ in Eq. \eqref{eqr11} is equal to $t_1\psi(2) + t_1\psi(0) + t_{-n}\psi(1-n)$ if $\psi(0)=\psi(1-n)=0$]. Substituting the expression for $\psi(x)$ in Eq. \eqref{psix} into these $n+1$ constraints results in a homogenous system of $n+1$ linear equations in the $n+1$ unknown $c_j$s. The eigenenergies of a chain with any finite value of $N$ can then be solved for exactly by finding the values of $E$ at which the determinant of this system of linear equations is zero.  

The GBZ approach provides a simpler approach for obtaining the loci of the eigenenergies on the complex energy plane in the thermodynamic limit $N\rightarrow\infty$ compared to computing the zeros of the determinant explicitly. The key idea in this is that a certain pair of the $\beta$ values is required to have the same moduli so that the boundary conditions can be satisfied at both ends of the chain simultaneously, as explained in the following. To facilitate the explanation, we shift the $x$ position labels of the chain from $x=1$ -- $x=N$ to $x=-(N-1)/2$ -- $x=(N-1)/2$. The boundary conditions then become $\psi(-(N-1)/2-n) = \psi((N-1)/2-n+1) = ...  = \psi(-(N-1)/2-1)=0$ and $\psi((N-1)/2+1)=0$. 

Consider first the boundary condition at $x=(N-1)/2+1$. As $N \rightarrow \infty$, the absolute value of $\beta_j^{(N-1)/2+1}$ for the smaller $\beta_j$ values with $j=1,..,n-1$ become negligibly small compared to those of $\beta_n$ and $\beta_{n+1}$. The corresponding $c_j\beta_j^{(N-1)/2+1}, j = 1, ..., n-1$ terms in Eq. \eqref{psix} can then be approximated to zero and we have
\begin{equation}
    \psi((N-1)/2+1) \approx c_{n}\beta_n^{(N-1)/2+1} + c_{n+1}\beta_{n+1}^{(N-1)/2+1}. \label{psiright}
\end{equation}

Note that we cannot approximate the $c_n\beta_n^{(N-1)/2+1}$ term in Eq. \eqref{psiright} to 0 because otherwise, the $c_{n+1}\beta_{n+1}^{(N-1)/2+1}$ cannot be cancelled off to make $\psi((N-1)/2+1)$ zero. 

Consider next the $n$ boundary conditions at the left end of the chain $\psi(x)=0, x=-(N-2)/2-n, ..., -(N-1)/2-1$. To guarantee the existence of a solution for these $n$ equations, we need all $n+1$ terms in Eq. \eqref{psix} to be of approximately the same order of magnitude at these values of $x$ so that there are more non-negligible free variables (i.e., the $n+1$ $c_j$s) than constraints (the $n$ boundary conditions). Now considering all the $n+1$ boundary conditions at the left and right ends collectively, we note that there is a requirement for
\begin{equation}
    |c_{n}\beta_n^x| \simeq |c_{n+1}\beta_{n+1}^x| \label{bcBothEnds}
\end{equation}
at both negative values of $x$ at $x=-(N-2)/2-n, ..., -(N-1)/2-1$ and at a positive value of $x$ at $x=(N-1)/2+1$. Eq. \eqref{bcBothEnds} can hold at both the negative and positive values of $x$ as $N\rightarrow\infty$ only when $|\beta_n|=|\beta_{n+1}|$; otherwise, if $|\beta_{n+1}|$ is slightly larger than $|\beta_n|$, the left side of Eq. \eqref{bcBothEnds} will become exponentially smaller than the right side at $x=(N-1)/2+1$, and exponentially larger than the right at $x=-(N-2)/2-n, ..., -(N-1)/2-1$ as $N\rightarrow\infty$ for any finite values of $c_n$ and $c_{n+1}$. The loci of the OBC energy eigenvalues at large values of $N$ therefore approaches the loci of $E$ at which $|\beta_n|=|\beta_{n+1}|$, which gives the GBZ. (This condition differs from the usual criteria that it is the moduli of the middle pair of $|\beta|$ that needs to have the same value rather than the largest pair of $|\beta|$ values here because the former applies only for systems at which the furthest coupling to the left and right have the same distances, whereas the long-range coupling here is unidirectional. )

We illustrate the application of the GBZ approach through the example of the system with second-order unidirectional coupling  in Eq. \eqref{eqr11}. The solutions for $\beta$ of Eq. \eqref{Epbc} at $n=2$ are then given by 
\begin{eqnarray}
\beta_1= \frac{E}{3t_1}-\frac{2^{\frac{1}{2}} \lambda_2}{3t_1 \lambda}+ \frac{\lambda}{3 \times 2^{\frac{1}{3}} t_1}, \label{n2b1} \\
 \beta_2= \frac{E}{3t_1}+ \frac{(1+\sqrt{3}i) \lambda_2}{3 \times 2^{\frac{2}{3}} t_1 \lambda} -\frac{(1-\sqrt{3}i) \lambda}{6 \times 2^{\frac{1}{3}} t_1} \label{n2b2},\\
\beta_3= \frac{E}{3t_1}+ \frac{(1-\sqrt{3}i) \lambda_2}{3 \times 2^{\frac{2}{3}} t_1 \lambda} -\frac{(1+\sqrt{3}i) \lambda}{6 \times 2^{\frac{1}{3}} t_1} \label{n2b3},
 \end{eqnarray}
where $\lambda_1 = 2 E^3-27 t_1^2 t_{-2}^2-9E t_1 t_{-1}$, $\lambda_2=3t_1 t_{-1}-E^2$ and $\lambda=\left( \lambda_1+ \sqrt{4 \lambda_2^3+\lambda_1^2} \right)^{1/3}$. Following the arguments above, the OBC eigenspectrum is given by the loci of $E$ where $|\beta_2|=|\beta_3|$. Although the loci of $E$ that satisfies this requirement is obviously too complicated to solve for analytically, it can be appreciated from the presence of the complex coefficients in Eq. \eqref{n2b2} and \eqref{n2b3} that the solutions for $E$ are, in general, complex and not purely real. Moreover, the common value of $|\beta_2|$ and $|\beta_3|$ is, in general, not necessarily 1. This results in an exponential localization of the wavefunction, i.e., the NHSE via Eq. \eqref{psix}. 

For comparison, we also derive the OBC eigenenergy spectrum of the HN chain without long-range coupling (i.e., $t_n=0$) using the GBZ approach. In this case, the two values of $\beta$ are given by 
\begin{equation}
    \beta_\pm = \frac{ E \pm \sqrt{E^2 - 4 t_1t_{-1}}}{2 t_{1}}, \label{b2pm}
\end{equation}
and it is required that $|\beta_+| = |\beta_-|$ on the GBZ. A key difference between Eqs. \eqref{b2pm}, for which there is no long-range coupling, and \eqref{n2b2} and \eqref{n2b3}, for which there is a second-order long-range coupling, is that there are no complex coefficients in the former. This opens the possiblity for the solutions of $E$ in $|\beta_+| = |\beta_-|$ to be purely real in the HN chain rather than complex. Indeed,  $|\beta_+| = |\beta_-|$ holds when the $\beta_\pm$s form a complex conjugate pair. This occurs when the term in the square root, viz. $E^2 - 4 t_1t_{-1}$, is negative. The OBC spectrum of the HN chain in the thermodynamic limit is hence given by $|E| < 2\sqrt{t_1 t_{-1}}$, which matches the OBC spectrum obtained using the imaginary gauge approach.


%

\end{document}